\documentclass[article,twoside]{revtex4}
\usepackage{color}
 \usepackage{amsbsy}
  \usepackage{bm}
 \usepackage{amsfonts}
\usepackage{amssymb}
\usepackage{amsmath}
\usepackage{graphics}
\usepackage{graphicx}
\begin{document}
\title{Effects of equilibrium coexisting phases in the first-order chiral
 transition within the Linear sigma model with quarks}

\author{R. M. Aguirre}
\affiliation{Departamento de Matematica, Universidad Nacional de
La Plata\\ and Instituto de Fisica La Plata,  CONICET\\ Argentina}

\begin{abstract}
The first order chiral phase transition for quark matter with
flavor imbalance is studied using the Linear sigma model with
quarks, also known as Quark-meson model. Special attention is paid
to the role of the scalar isovector meson. The general consensus
presently is that the chiral transition changes from a smooth
crossover to first-order at low temperatures.  This transition is
assumed to be discontinuous, with unstable or metastable
intermediate states. However, if multiple charges are
simultaneously conserved the system could undergo a continuous
change through a coexistence of equilibrium states. Under such
assumption the bulk properties are analyzed and several remarkable
effects for the speed of sound and the susceptibilities are
stressed.

\end{abstract}

\maketitle

\section{INTRODUCTION}
The gross features of the QCD phase diagram have been deduced long
time ago, however the precise details are still pending. Intensive
investigations on strong interacting matter under extreme
conditions of density and temperature have scrutinized the
different regimes, providing information on new aspects.
Experiments on heavy ion collisions, for instance, have provided a
lot of evidence to understand the high temperature and low matter
concentration domain. On the other hand, the observation of
compact astronomical objects permanently produce new data which is
elaborated to understand the low temperature and medium-high
density regime. Additional precisions have been obtained from the
lattice simulation of the fundamental theory (LQCD)
\cite{ALFORDK,SON,NISHIDA,EJIRI}. This method is particularly
appropriate to study the vacuum properties and finite temperature
effects, although finite density systems are not accessible due to
the well known problem of the negative sign. To avoid this
technical difficulty the study of matter at zero baryon density
but with isospin imbalance, has been proposed \cite{ALFORDK}. The
conservation of the isospin charge, with the introduction of the
corresponding chemical potential $\mu_I$, eliminates the problem
of the determinant.

Complementarily, the regime of medium-high baryon densities and
low temperatures is commonly analyzed using effective models since
first principle calculations, as in lattice simulations, are not
feasible. Several theoretical models highlighting different
aspects of the strong interaction have been used. The compilation
of their predictions has produced a complex picture of the phase
diagram, with diverse phase transitions, condensates, collective
states, etc.

Due to the property of asymptotic freedom of the QCD theory, it
has been hypothesized that an state of deconfined quarks and
gluons could be obtained under extreme conditions.  This
deconfinement phase transition has particularly been analyzed
\cite{ARYAL,PRAKASH} since it could have clear manifestations in
heavy ion collision events or, as pointed out more recently, it
could leave its imprint in the gravitational waves  coming from
the collapse of companion neutron stars.

The recovery of the chiral symmetry is also expected to occur
through a phase transition following a pattern similar to the
deconfinement case. A change of character from first-order at low
temperatures to second order at higher ones is expected. If the
quarks have a non-zero mass, the second order transition becomes a
soft crossover. In any case the line of transition points in the
$(\mu_B,T)$ plane exhibits a critical endpoint (CEP) separating
different regimes.

General features of the chiral transition in quark matter with
isospin imbalance have been investigated in
\cite{TOUBLAN,FRANK,BARDUCCI,BARDUCCI1,LOEWE,HE,HE1,ZHANG,MUKHERJEE,KOVACS0,MU,ABUKI,XIA,STIELE,LIU,AVANCINI,LOPES,ADHIKARI,COSTA,CHU,AYALA,AYALA0},
The coherence of the theoretical predictions with the LQCD results
at zero baryon number has been particularly analyzed in
\cite{LOEWE,AVANCINI,LOPES,ADHIKARI,AYALA0}.

It has been argued that in the limit of zero baryonic density
($\mu_B=0$) at constant isospin fraction, the associated chemical
potential $\mu_I$ can be treated as a gauge field \cite{SON}.
Under this hypothesis the pion field could experience a
condensation along the chiral transition. However, it has been
pointed out that a system with $\mu_I
> \mu_B$ could be unstable by weak decay.

The fluctuations of conserved charges are good markers of the
phase transitions, being related to the corresponding
thermodynamical susceptibilities by the fluctuation dissipation
theorem \cite{LAHIRI}. For this reason the susceptibilities
associated with the baryonic number or electric  charge, have been
focused by the LQCD. Furthermore, they provide local information
of the equation of state.

The aim of this paper is the study of the phase diagram near the
chiral transition, by taking the quark number density and the
temperature as thermodynamical variables and considering the
conservation of the isospin asymmetry. The Linear Sigma Model with
quarks (LSMq) is specially suited for this purpose, as it was
originally conceived to reproduce the low energy phenomenology of
QCD. The weak interaction is not considered here, as it is a
common practice \cite{SON}. Therefore, the $u$ and $d$ quark
flavors are stable degrees of freedom. Furthermore, an scalar
isovector meson $\zeta$ is introduced and treated on the same foot
as the standard $\sigma, \pi$ ones. This field propagates the
flavour interaction between quarks. An enlarged family of mesons
within the LSMq have been considered previously
\cite{PARGANLIJA,PARGANLIJA1,KOVACS}, and particularly the role of
the $\zeta$ meson
\cite{GASIOROWICZ,METZGER,LENAGHAN,SCHAEFER,ABUKI,AYALA1}. In
order to stress the results on thermodynamical instabilities, the
simplicity of the theoretical formulation is preferred here.
Hence, only the meson fields $\sigma,\, \pi, \,\zeta$ and the
 Hartree approach are considered here, to take
advantage of the simplicity and versatility of the model.
Notwithstanding a more sophisticated approach is in progress.

The rest of this work is organized as follows, the mean field
approach to the LSMq, also known as Quark Meson model, is
presented in the next section. The results for the thermodynamics
of the system is discussed in Sec. \ref{RESULTS}, and the final
conclusions are drawn in Sec. \ref{SUMMARY}.
\section{THE MODEL}

The simplest version of the LSMq, with the only addition of the
$\zeta$ meson, can be written as \cite{AYALA1}
\begin{eqnarray}{\cal
L}&=&\,\bar{\Psi}\left( i \not \!
\partial+g \,\Phi \right)\Psi+\frac{1}{2}\left(\partial_\mu \sigma
\partial^\mu \sigma+ \partial_\mu \bm{\pi} \cdot
\partial^\mu \bm{\pi}+\partial_\mu \bm{\zeta} \cdot
\partial^\mu \bm{\zeta}\right)+\frac{1}{4}C_0 \text{Tr}
\left(\Phi^\dag \Phi\right)\nonumber \\
&&-\frac{1}{16} C_3 \left[\text{Tr} \left(\Phi^\dag
\Phi\right)\right]^2-\frac{1}{8} C_2 \text{Tr} \left(\Phi^\dag
\Phi\right)^2-h\, \sigma \label{LAGRANGE}
\end{eqnarray}
the two quark flavors are collected in the bi-spinor
$\Psi=\left(\psi_u\;\psi_d\right)^t$, the meson $\sigma$ and the
iso-multiplets scalar $\bm{\zeta}$, and pseudo-scalar $\bm{\pi}$,
constitute $\Phi=\sigma+i\,\gamma_5 \bm{\pi}\cdot
\bm{\tau}+\bm{\zeta}\cdot \bm{\tau}$. Flavor degeneracy in the
quark  coupling is assumed and a symmetry breaking term, linear in
$\sigma$  has been included.

The scalar meson fields can be decomposed as the sum of a thermal
expectation value and its fluctuation $\sigma=s+\delta \sigma$,
$\zeta_a=z\,\delta_{a3}+\delta \zeta_a$. The quark fields acquire
finite masses at the tree level, as can be easily seen after the
expansion in the meson fluctuations and examining terms
proportional to $\bar{\psi}_j \psi_j$. Thus, the effective quark
mass $m_j=g(s+I_j \, z)$ can be identified, where $I_j=1\, (-1)$
for $j=u\, (d)$. By the same procedure, but collecting quadratic
terms in the meson fluctuations $\delta \sigma^2,\, \delta
\zeta_a^2,\, \pi_a^2$, one can identify the meson masses
\cite{LENAGHAN}
\begin{equation}M_\sigma^2=-C_0+3 C_3
\,s^2+(C_3+2 C_2) \,z^2, \;M_{\pi\,a}^2=-C_0+C_3 (s^2+z^2),\;
M_{\zeta\,a}^2=-C_0+C_3\, n_a\,z^2+(C_3+2 C_2) \,s^2
\label{MesonMass}\end{equation}
with $n_a=1\, (3)$ for $a=1,2\,(3)$.

The mean field approach (MFA) is obtained when the meson
fluctuations are completely omitted, while its explicit
consideration can be arranged as higher order corrections
\cite{DOLAN,PETROPOULOS,CORNWALL}. Here I adopt the Hartree
approach, which consist of the MFA complemented with the vacuum
contribution from the Dirac sea of quarks. The MFA is the standard
approach in quark matter calculations using the effective
Nambu-Jona Lasinio model \cite{SASAKI,SASAKI1,SASAKI2}, and it has
been used in the case of the LSMq model too
\cite{STOKIC,SCHAEFER}. An equivalent treatment is usual in
relativistic hadronic models, and it is applied to study phase
transitions as the nuclear liquid-gas, or the quark deconfinement
in high density matter \cite{GLENDENNING}.

Within the Hartree scheme the grand potential per unit volume of
uniform quark matter can be written as
\begin{eqnarray}
\omega(T,\mu)&=&\omega_{\text{vac}}-\frac{N_c}{\beta
\pi^2}\sum_{j=u,d}\int_0^\infty
dp\,p^2\left[\text{log}\left(1+e^{-\beta(E_{p j}-\mu_j)}\right)
+\text{log}\left(1+e^{-\beta(E_{p j}+\mu_j)}\right)\right]
\label{GRAND P}
\end{eqnarray}
where
\begin{equation} E_{p j}=\sqrt{p^2+g^2(s+I_j z)^2},\;\;
I_j=1,-1\; \text{for} \,j=u,d\label{QSpectrum}\end{equation}
The vacuum term is divergent, but a finite contribution can be
extracted by dimensional regularization followed by an appropriate
subtraction scheme,  as shown in the Appendix
\begin{equation}\omega_{\text{vac}}=\frac{N_c}{8
\pi^2}\sum_{j=u,d}\left[m_j^4 \,\text{log}\left(\frac{m_j}{m_0}
\right)+\frac{m_0^4-m_j^4}{4}\right]  \label{GP VAC}\end{equation}
Here $m_0$ stands for the degenerate mass physical value.

As usual,the definition of the mean values $s,\, z$ are given by
the stationary conditions
\[ 0=\frac{\partial \omega}{\partial s},\;\;\;\;\;\; 0=\frac{\partial \omega}{\partial z},\]
which lead to the equations
\begin{equation}
0=(-C_0+C_3 s^2+C_4 z^2)\,s-h-\frac{g
N_c}{\pi^2}\sum_{j=u,d}m_j\left[\frac{m_j^2}{2}\,\text{log}\left(\frac{m_j}{m_0}
\right)-\int_0^\infty \frac{dp\; p^2}{E_j}\left(n_{F j}+\bar{n}_{F
j}\right) \right],\label{S MFA}
\end{equation}
\begin{equation}
0=(-C_0+C_3 z^2+C_4 s^2)\,z-\frac{g N_c}{\pi^2}\sum_{j=u,d}I_j
\,m_j\left[\frac{m_j^2}{2}\,\text{log}\left(\frac{m_j}{m_0}
\right)-\int_0^\infty \frac{dp\; p^2}{E_j}\left(n_{F j}+\bar{n}_{F
j}\right) \right],\label{Z MFA}
\end{equation}
respectively. In Eqs.(\ref{S MFA}, \ref{Z MFA}) the equilibrium
Fermi distribution functions $n_{F j}(T,\mu_j)$ for quarks and
$\bar{n}_{F j}(T,\mu_j)$ for antiquarks have been introduced.

The entropy of the system is given by the thermodynamical
definition ${\cal S}=S/V=-\partial\,\omega/\partial T$ as
\begin{eqnarray}
{\cal
S}(T,\mu)&=&-\frac{\omega-\omega_{\text{vac}}}{T}+\frac{N_c}{
\pi^2 T}\sum_{j=u,d}\int_0^\infty dp\,p^2\left[(E_j-\mu_j)\, n_{F
j}+(E_j+\mu_j)\, \bar{n}_{F j}\right] .\label{ENTROPY}
\end{eqnarray}
Finally, the energy density is evaluated by the Legendre transform
${\cal E}=\omega+ T \,{\cal S}+\sum_{j=u,d}\mu_j\,n_j$, using the
quark number density
\begin{equation}n_j=\frac{N_c}{\pi^2}\int_0^\infty dp\,p^2\left( n_{F
j}- \bar{n}_{F j}\right).\label{NumberDens}\end{equation}

In the following the evolution of matter at fixed isospin fraction
$x=(n_d-n_u)/(n_u+n_d)$ is considered. Some characteristic
quantities  of the thermodynamical evolution of the system are the
specific heat capacities $c_V=T \left(\partial {\cal S}/\partial
T\right)_n$, $c_P=T \left(\partial {\cal S}/\partial
T\right)_{n,P}$, the isothermal speed of sound
$v_T=-\left(\partial \omega/\partial {\cal E}\right)_{T x}$ and
the second  order susceptibilities $\chi$
\cite{GAVAI0,FERRONI,RATTI,SASAKI,SASAKI1,SASAKI2,STOKIC,ALMASI,SKOKOV,SKOKOV1,SKOKOV2}.
The speed of sound is a useful indicator of the emergence of new
degrees of freedom. Meanwhile the susceptibilities manifest the
dynamical fluctuations characterizing the bulk (semi-classical)
behavior of the system \cite{ASAKAWA,GAVAI,DATTA}. Fluctuations
are closely related to phase transitions, in particular those
related to conserved charges. For such reason the susceptibilities
associated with conserved charges have been focused within the
LQCD \cite{RATTI,GAVAI,DATTA}.

In the following the susceptibilities associated with the quark
numbers
\[ \chi_B=\left(\frac{\partial n_B}{\partial \mu_B}\right)_{\mu_3},\;\;
\chi_3=\left(\frac{\partial n_3}{\partial
\mu_3}\right)_{\mu_B},\;\;\chi_{B x}=\left(\frac{\partial
n_B}{\partial \mu_B}\right)_x,\;\;\chi_{3 x}=\left(\frac{\partial
n_3}{\partial \mu_3}\right)_x\]
 will be considered. Here the
densities for the baryonic number $n_B=(n_d+n_u)/3$, and the
isospin number $n_3=n_d-n_u$ have been introduced, together with
the chemical potentials for the baryonic $\mu_B=3(\mu_d+\mu_u)/2$
and the isospin $\mu_3=\mu_d-\mu_u$ charges.

In a preliminary study \cite{AGUIRRE} some alternatives to the
present approach were considered. For instance in the one loop
approximation,  as described in \cite{AYALA0}, the mesons behave
as free particles with effective masses given by Eq.
(\ref{MesonMass}). The grand potential receives additional
contributions
\[
\omega_{\text{1 loop}}(T,\mu)=\omega(T,\mu)+\frac{1}{2 \beta
\pi^2}\sum_{\alpha=\sigma,\pi_a,\zeta_a}\left\{\int_0^\infty
dp\,p^2\,\text{log}\left(1-e^{-\beta\,E_\alpha}\right)-\frac{\beta}{16}
\left[M_\alpha^4 \,\text{log}\left(\frac{M_\alpha}{M_{0 \alpha}}
\right)+\frac{M_{0 \alpha}^4-M_\alpha^4}{4}\right]\right\}
\]
This scheme breaks down at nonzero density because the pion mass
becomes imaginary. In flavor symmetric matter it is found that
$M_\pi^2<0$ as soon as $s/f_{\pi}<0.9$, which happens at
relatively low density. The problem of the imaginary meson mass at
finite temperature and zero density is known from long time ago
\cite{GOLDBERG}, and it has motivated alternatives to the ordinary
loop expansion. The CJT formalism \cite{CORNWALL,CAMELIA}, for
instance, is efficient to treat this failure at zero density
\cite{RODER}. The grand potential is considered a functional of
the mean values $s, z$ and the full meson propagators, for this
reason the effective masses are given by coupled self-consistent
equations \cite{CORNWALL,CAMELIA,PETROPOULOS}. An attempt to use
the CJT method within the two-flavor LSMq to describe dense
systems produce similar results, but in this case both the pion
and the $a_0$ meson masses become imaginary.

Thus I restrict here to the Hartree approach, expecting that it
serves as a guide to develop an approach to dense matter including
the effect of meson fluctuations.

\section{RESULTS AND DISCUSSION}\label{RESULTS}

The parameters of the model (\ref{LAGRANGE}) are fixed in order to
reproduce the physical masses of the mesons in vacuum
\[C_0=\frac{M_{\sigma 0}^2-3 M_{\pi 0}^2}{2},\; C_2=\frac{M_{\zeta 0}^2-3 M_{\pi 0}^2}{2 f_\pi^2},
C_3=\frac{M_{\sigma 0}^2- M_{\pi 0}^2}{2 f_\pi^2},\]
following the standards of the model it is assumed that $s=f_\pi$
in the same conditions. Furthermore the values $M_{\sigma 0}=500$
MeV, $M_{\pi 0}=138$ MeV, and $M_{\zeta 0}=984$ MeV have been
used. In particular the value of $M_{\zeta 0}$ arises from the
identification of the $\zeta$ meson with the $a_0$(980), since
this is the lightest manifestation in the hadronic sector of an
scalar isovector. Finally, the coupling constant $g$ is related to
the quark constituent mass through $m_{q 0}=g\,f_\pi$ and it will
be discussed in the next step.

The chiral phase transition in homogeneous quark matter, keeping
constant the isospin fraction, is analyzed using the model thus
defined. The relation between chiral breakdown and deconfinement
is not clear yet, and for the sake  of simplicity it is not
treated here. However, this subject has been extensively studied,
specially with the aid of a phenomenological Polyakov potential.
\cite{MUKHERJEE,RATTI,SKOKOV1,SKOKOV2,GHOSH,LIU}. Furthermore,
neither the pion condensation nor the superconducting quark phase
will be taken into account here since the conditions considered in
this work are below the threshold of these phenomena. For
instance, the isospin chemical potential verifies $\mu_3<M_{\pi
0}$. In fact, at fixed $x$ the chemical potential $\mu_3$
increases with $n$, and for the highest density considered $n/n_0=
5$  it is found $\mu_3/M_{\pi 0}<0.45 (0.95)$ for $x=1/3 (2/3)$
and $T \leq 20$ MeV.

\begin{figure}
    \centering
    \includegraphics[height=0.4\textheight] {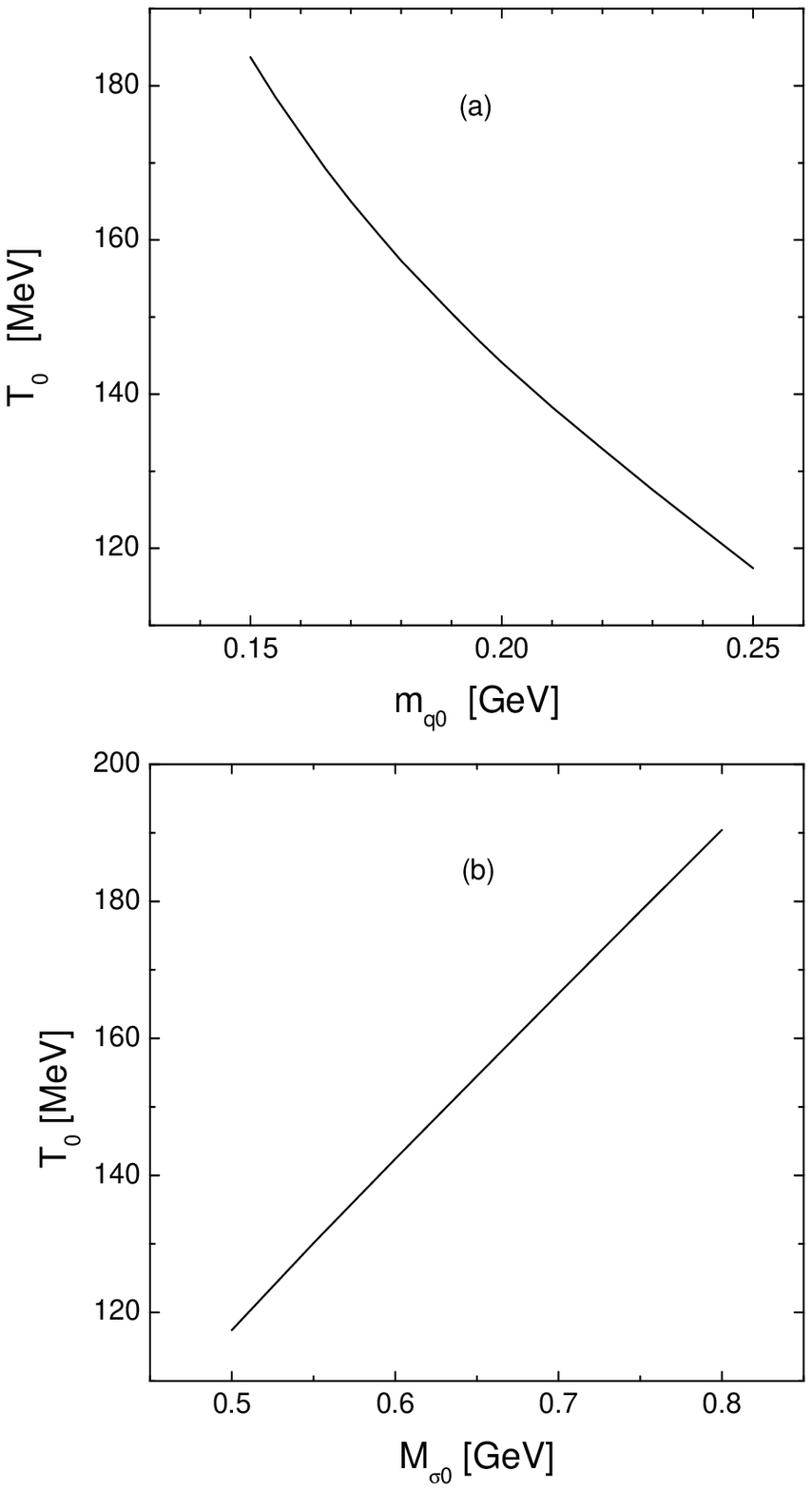}
    \caption{The critical temperature in terms of the quark mass (a), and its dependence on
    the model parameter $M_{\sigma 0}$ (b).  The quark mass in vacuum is related to the
    coupling constant by $m_{q 0}=g\, f_\pi$.  }
    \label{Fig Tcritical}%
\end{figure}

\begin{figure}
    \centering
    \includegraphics[height=0.4\textheight]{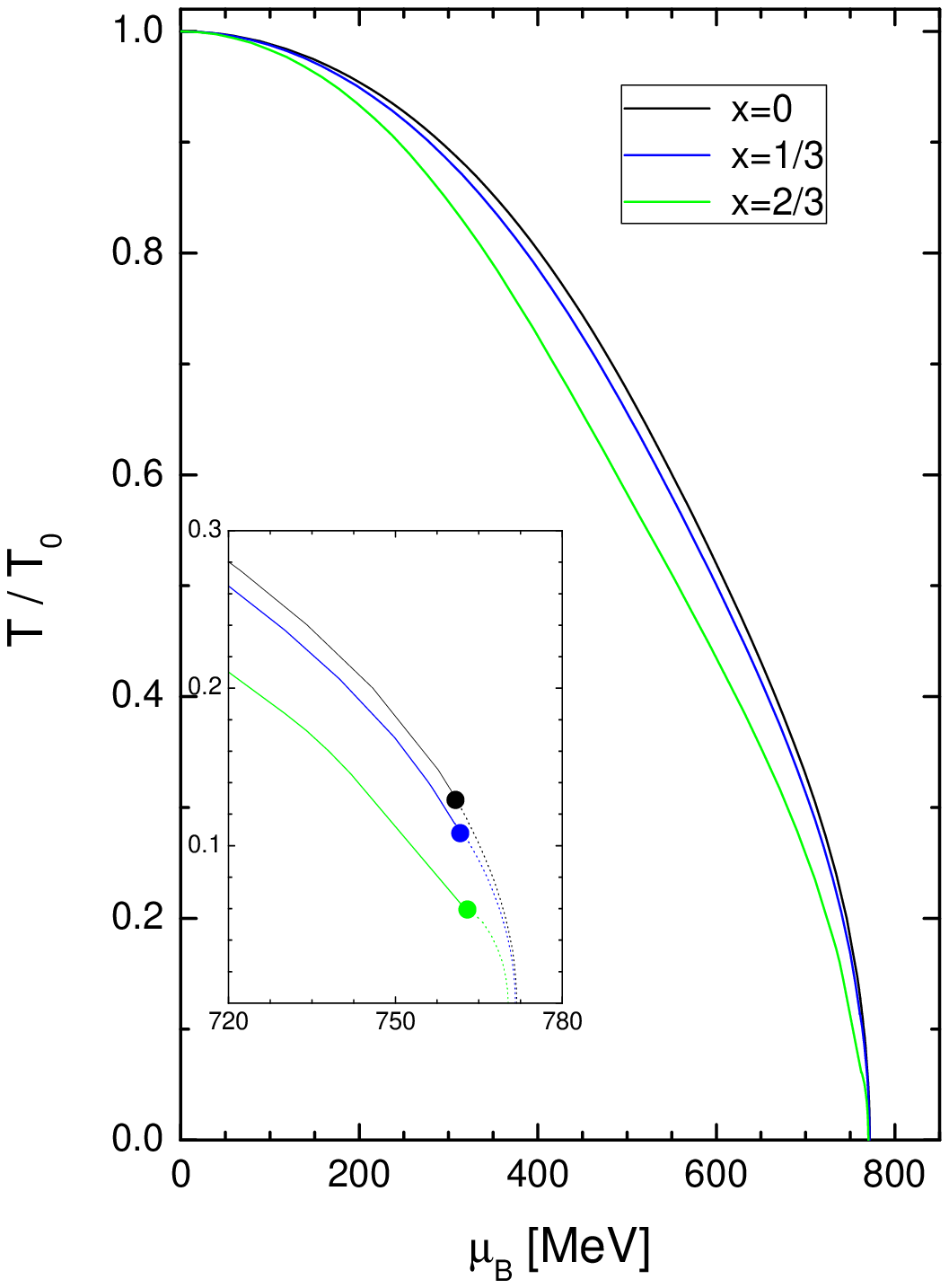}
    \vspace{-1cm}\caption{The transition temperature in terms of the baryon number chemical potential for several global
    isospin parameters $x$. The insertion shows details of the first order transition (dotted lines),
    the CEPs (full circles) and the crossover transition (solid and dashed lines).
    Within the ECR the Eq.(\ref{x Binodal}) applies.  }
    \label{Fig Critical}%
\end{figure}

There is currently a general consensus that the chiral transition
is a continuous crossover for low densities and high temperatures
if  massive quarks are assumed. Since there is no unambiguous
definition of the transition temperature in a crossover, it is
defined here as the temperature of the inflection point of the
chiral order parameter \cite{PAWLOWSKI}, i.e. where $s^2+z^2$
reach its minimum value. In the present approach the critical
temperature $T_0$ is a decreasing function of $g$, as shown in
Fig. \ref{Fig Tcritical}a, where a range $150$ MeV $\leq m_{q 0}
\leq 250$ MeV has been considered. As can be seen, the
compatibility with LQCD estimations \cite{EJIRI} is improved
adopting low values of $m_{q 0}$. However, for such range of
constituent quark masses the expected change of regime, through an
intermediate CEP, is frustrated. Taking $m_{q 0}=250$ MeV the
value $T_0 \simeq 120$ MeV is obtained, which is out the range
$160-180$ MeV predicted by LQCD. This could be an effect of the
parametrization used. The correlation between $M_{\sigma 0}$ and
the critical temperature has been studied in \cite{SCHAEFER,RAI},
in different versions of the LSMq. An increasing dependence of
$T_0$ with the meson mass was found there, as well as a decreasing
trend of the CEP temperature. This relation, shown in Fig.
\ref{Fig Tcritical}b, leads to the conclusion that an increment of
a few tenths of GeV for $M_{\sigma 0}$ will improve the output for
the critical temperature. In any case, a better fit could be
obtained if higher order corrections in the meson sector are taken
into account \cite{BILIC,PETROPOULOS,AYALA1}, or if the Polyakov
loop potential is included \cite{SCHAEFER2}. A value of compromise
$m_{q0}=250$ MeV, keeping $M_{\sigma 0}=0.5$ GeV, is adopted here
to fulfill the phenomenological requirements as best as possible.
Using the parametrization just described the phase diagram is
evaluated for several values of $x$, obtaining the result  shown
in Fig.\ref{Fig Critical}. The illustrative values selected for
the flavor asymmetry are as follows, $x=0$ corresponds to the
isospin symmetric case usually discussed in finite density
calculations, $x=1/3$ represents electrically neutral matter as
can be found in self-bound quark stars, and finally $x=2/3$ is an
extrapolation to the negatively charged case.
 There is a wide variety of predictions about the
position of the CEP, even using different parameterizations of the
same model. An analysis of this type of uncertainty was carried
out in \cite{BIGUET} using the NJL model, concluding that the
value of the chemical potential is a robust prediction of each
model, while the temperature is strongly dependent on the
parametrization used. In the present approach the location of the
CEP changes slightly with the flavor composition. Although its
chemical potential $\mu_B \simeq 770$ MeV is almost independent of
$x$, its temperature varies within $7-15$ MeV. Restricting the
comparison for the position $(\mu_B,T)$ of the CEP to results
obtained with the LSMq, one can find for instance (880,30) MeV in
\cite{AYALA1}, (840,85) MeV, and (860,30) MeV in \cite{SCHAEFER2},
and (900,30) MeV in \cite{RAI}. The first case considers two
flavors with two-loop mesonic corrections, the second case
corresponds to 2+1 flavors with or without Polyakov potential, and
the last case deals with two flavors and the octet of mesons, and
parameters fixed by a specific scheme. As a general conclusion it
can be stated that the present approach predicts a more reduced
extension of the broken chiral phase.  In addition the crossover
transition temperature decreases with $x$ for a given chemical
potential $\mu_B$.

\begin{figure}
    \centering
    \includegraphics[height=0.4\textheight]{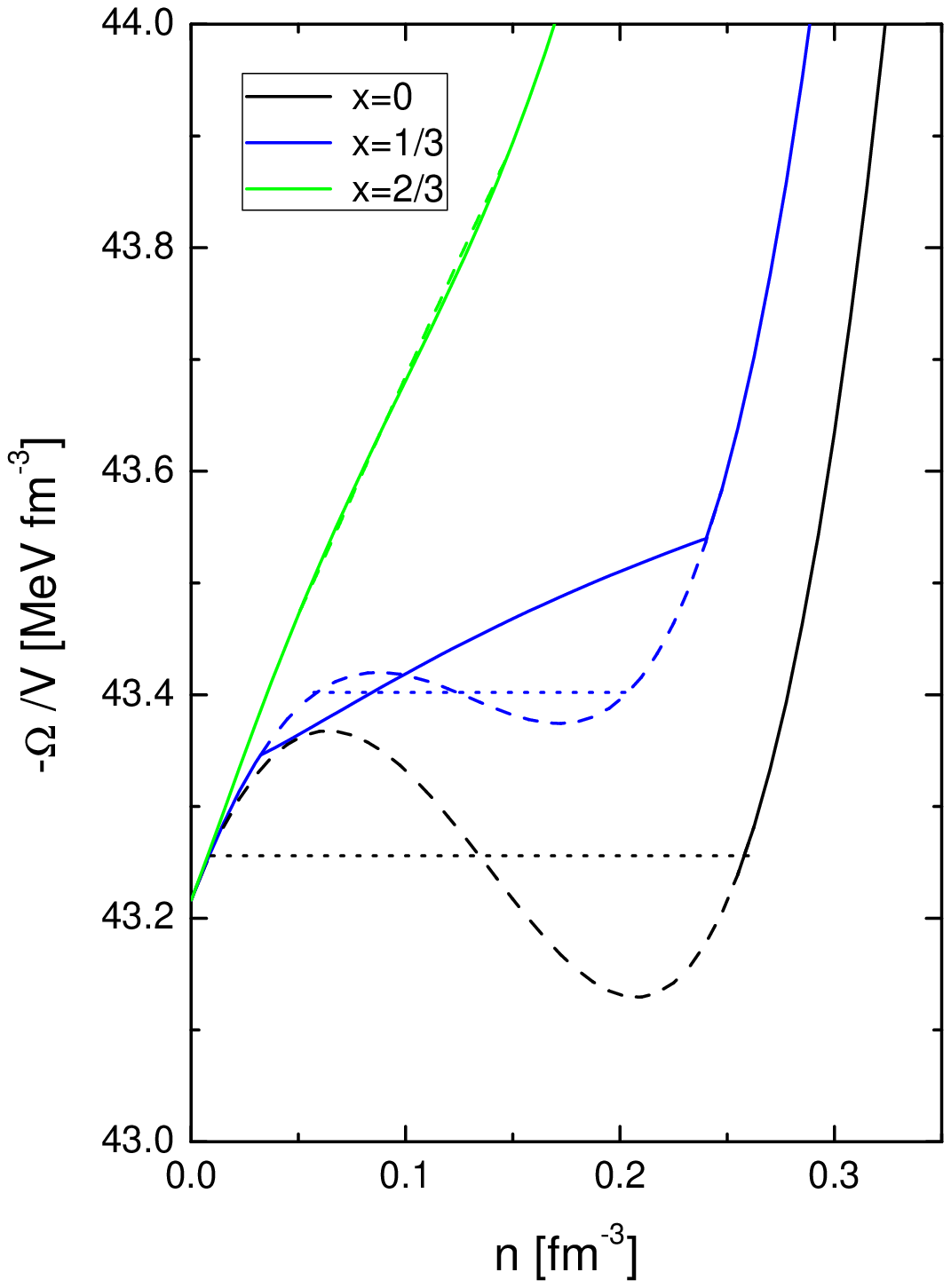}\vspace{-1cm}
    \caption{The isothermal pressure at $T=5$ MeV as function of the quark number for several
    flavor asymmetries $x$. Solid lines corresponds to the results
    including the ECR. The Maxwell construction for $x=0,\,1/3$ is represented by dotted lines,
    while the dashed lines include contributions from  unstable and meta-stable configurations.   }
    \label{Fig EoS}%
\end{figure}

Under certain conditions the thermodynamic potential $\omega$ does
not depend monotonically on the chemical potentials or,
equivalently, on the number densities. This fact is reflected by
the presence of local extrema in the pressure, as exemplified in
Fig.\ref{Fig EoS}.  The isotherms for $T=5$ MeV, corresponding to
different flavor compositions, exhibit some typical situations
which generate thermodynamical instabilities causing a first order
phase transition.  Depending on the value of the surface tension,
the instabilities are resolved as a discontinuous change or as a
continuous passage with an intermediate coexistence region. The
last case is feasible because there are multiple conserved charges
\cite{GLENDENNING,HEISELBERG}, and  both coexisting phases
correspond to the regime of broken chiral symmetry. This type of
transition has multiple manifestations, as in the nuclear
liquid-gas transition causing the spinodal fragmentation.
Assuming a low surface tension the system evolves by following the
Gibbs construction. It is obtained by the introduction of a
parameter $\lambda$ which represents the relative abundance of the
coexisting equilibrium states, named $a$ and $b$ in the following.
Hence, the equilibrium conditions for coexisting phases read
\begin{equation}T_a=T_b,\;\; \mu_B^a=\mu_B^b,\;\;
\mu_3^a=\mu_3^b,\label{GIBBS1}\end{equation}
\begin{equation}P_a(T_a,\mu_B^a,\mu_3^a)=P_b(T_b,\mu_B^b,\mu_3^b),
\label{GIBBS2}\end{equation}
while for every additive thermodynamical function one has
\[ n_B=\lambda\,n_B^a+(1-\lambda)\,n_B^b,\; {\cal S}=\lambda\,{\cal S}^a+(1-\lambda)\,{\cal S}^b,\;
{\cal E}=\lambda\,{\cal E}^a+(1-\lambda)\,{\cal E}^b, \;
\text{etc.}\]
 $0\leq \lambda \leq 1$.

The curves with full lines in Fig.\ref{Fig EoS} correspond to the
physical situation, including the coexistence of phases through
the Gibbs construction. The dotted lines represent the Maxwell
construction, whose extremes are located at the onset of the
metastable region. Finally, the dashed lines indicate the part of
the equation of state replaced by the Gibbs construction. For
$x=0$ the Gibbs and Maxwell construction coincide since there is
only one conserved charge, due to the assumption of degeneracy of
the flavors $u, \, d$. The coexistence of phases extends to the
$x=2/3$ case, although there is no thermodynamical instability,
and the replacement has no significant effects on the pressure.
For $x=1/3$ the parameter $\lambda$ increases monotonously,
exhausting its domain as the system evolves. In contrast, for
$x=2/3$, the parameter $\lambda$ grows from zero to a maximum
value, lesser than $1$, and then returns to zero. For this reason
the full and dashed lines for $x=2/3$ almost coincide . In this
situation, known as retrograde transition, the thermodynamical
fluctuations are still strong, the system explores its
neighborhood in the phase space, but does not find a more
preferable configuration.
\begin{figure}
    \centering
    \includegraphics[height=0.35\textheight]{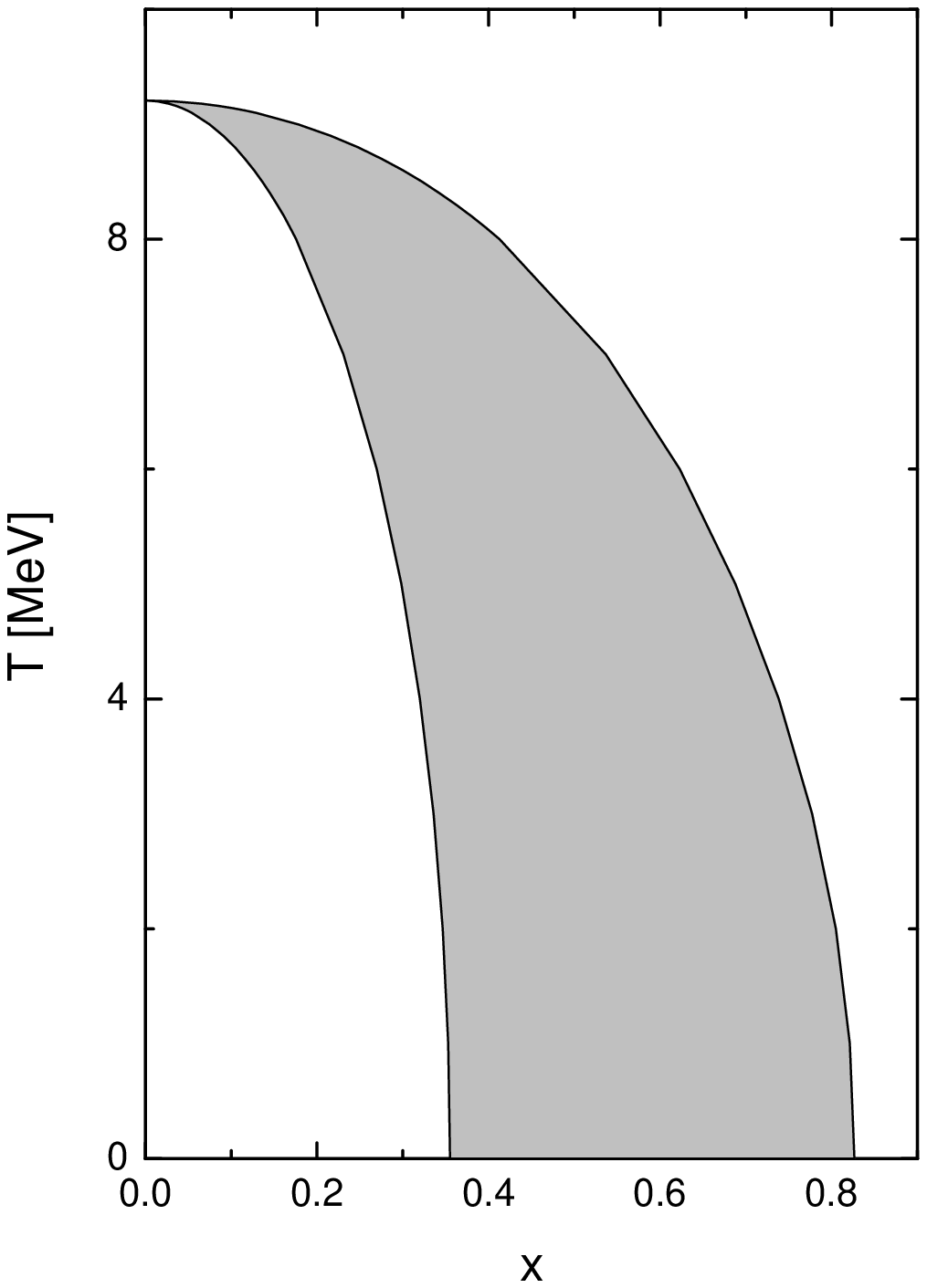}\vspace{-1cm}
    \caption{An isobar section of the equilibrium coexistence region in the $x-T$ plane corresponding to $P=45$ MeV fm$^{-3}$.}
    \label{Fig Binodal}%
\end{figure}
Within the instability region, the thermodynamical potential
$\omega(T,s,z,\mu_B,\mu_3)$ exhibits a non-monotonous behavior.
Thus, for a given temperature one can find two different sets of
variables $(s_a,z_a,\mu_{B a},\mu_{3 a})$ and $(s_b,z_b,\mu_{B
b},\mu_{3 b})$  which reproduce the same pressure, i.e. they
verify Eq.(\ref{GIBBS2}). If the corresponding states are
coexisting, then Eq. (\ref{GIBBS1}) must be imposed. However such
states could have very different mean field values $(s_a, z_a)$
and $(s_b, z_b)$. Using Eqs. (\ref{QSpectrum}) and
(\ref{NumberDens}) one can find $(n_{u a}, n_{d a})$ and $(n_{u
b}, n_{d b})$ or, equivalently, the quark number densities $n_a,
n_b$ and flavor asymmetries $x_a, x_b$ for both of these states.
In the following the evolution of quark matter at fixed global
isospin composition $x$ is described. Hence, this value must be
kept fixed even along the coexistence region. Thus, the constraint
\begin{equation}
x=\frac{\lambda\,x_a\,n_a+(1-\lambda)\,x_b\,n_b}{\lambda\,n_a+(1-\lambda)\, n_b}. \label{x Binodal}
\end{equation}
must be fulfilled for coexisting phases. Inverting this relation
one can find the value $\lambda$ corresponding to such pair of
coexisting states.

The collection of all these points, for the full range of
temperatures allowed, constitutes the Equilibrium Coexistence
Region (ECR). It is limited by the binodal surface and includes
the part of the phase diagram originally interpreted as unstable
or metastable. The ECR has been described in \cite{COSTA2,LiuZhou}
using the NJL model. To illustrate this subject an isobar section
of the ECR, corresponding to $P=43.5$ MeV fm$^{-3}$, is shown in
Fig. \ref{Fig Binodal}. It extends up to a maximum temperature
$T=9.2$ MeV, where only states with $x_a \simeq x_b \simeq 0$ are
involved. As a general trend it is found that the extension of
this region reduces, until it completely disappears when
\textit{i}) the pressure increases, or \textit{ii}) $g$ decreases.

A correct evaluation of the derivatives involved in the definition
of some bulk properties, such as heat capacity, speed of sound,
etc., must take account of the dependence of the parameter
$\lambda(\mu,T)$.
\begin{figure}[h]
    \centering
    \includegraphics[height=0.4\textheight]{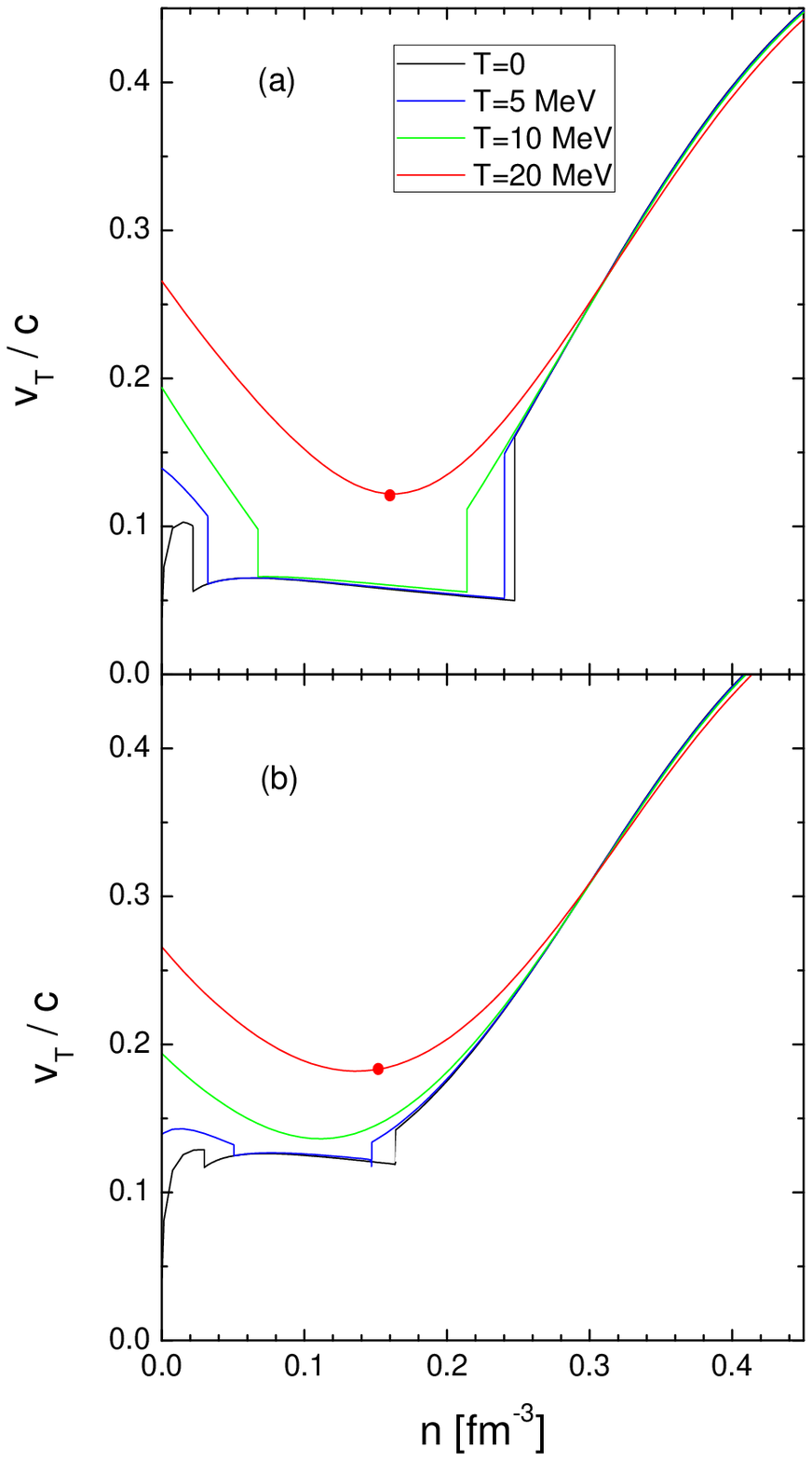}\vspace{-0.5cm}
    \caption{The isothermal speed of sound $v_T$, for configurations keeping $x$
    constant, as function of the quark number density for several temperatures and $x=1/3$ (a), and $x=2/3$ (b).
    The reference density is $n_0=0.15$ fm$^{-3}$.}
    \label{Fig Speed}%
\end{figure}

The speed of sound is an important ingredient in different
physical situations. The isothermal $v_T$ as well as the adiabatic
$v_S$ speeds enter in the definition of the tidal deformability of
compact stars \cite{ZACCHI}, which manifests in their non-radial
oscillations. For this reason it can indirectly be estimated from
the analysis of the gravitational waves coming from the collapse
of binary systems. This issue is under permanent attention due to
the evolution of observational methods and technologies.
Furthermore, it has been proposed that the high temperature
behavior of $v_S$ can be inferred from the experimental data on
ultrarelativistic heavy ion collisions \cite{GARDIM,MU5}. This is
in part the strong motivation for the recent studies on the speed
of sound in quark matter using effective models
\cite{AYALA,AYALA0,GHOSH,HE,HE1,LIU,DOSSOW}. Some of them explore
the limit $\mu_B \rightarrow 0$ in order to contrast with LQCD
results, and others extend to finite densities as for example the
recent study of the relation between the speed of sound and the
CEP \cite{DOSSOW}.

The isothermal speed of sound, along curves at constant $x$, is
shown in Fig.\ref{Fig Speed} as a function of the quark number
density for two isospin compositions and a range of $n$ and $T$
compatible with the first-order transition. For $x=1/3$ (upper
panel) and a temperature above the CEP, $T=20$ MeV, the speed
$v_T$ has a non-monotonic trend with its minimum located
approximately at the transition point (filled circle). For lower
temperatures the system experiences the coexistence of phases,
reflected by a discontinuous drop to a lower plateau. The vertical
segments at the discontinuities have the sole purpose of
facilitating the interpretation of the curves. Along the ECR, the
speed of sound has a mild variation with the particle number, and
it is approximately independent of the temperature. Furthermore,
the extension of the ECR is reduced as $T$ grows, a fact that
manifests by the shrinking of the plateau. A similar description
applies to $x=2/3$ (Fig. \ref{Fig Speed}b), but in such case only
the isotherms corresponding to $T=0, \text{and}\; 5$ MeV undergo a
first order transition.

\begin{figure}[b]
    \centering
    \includegraphics[height=0.4\textheight]{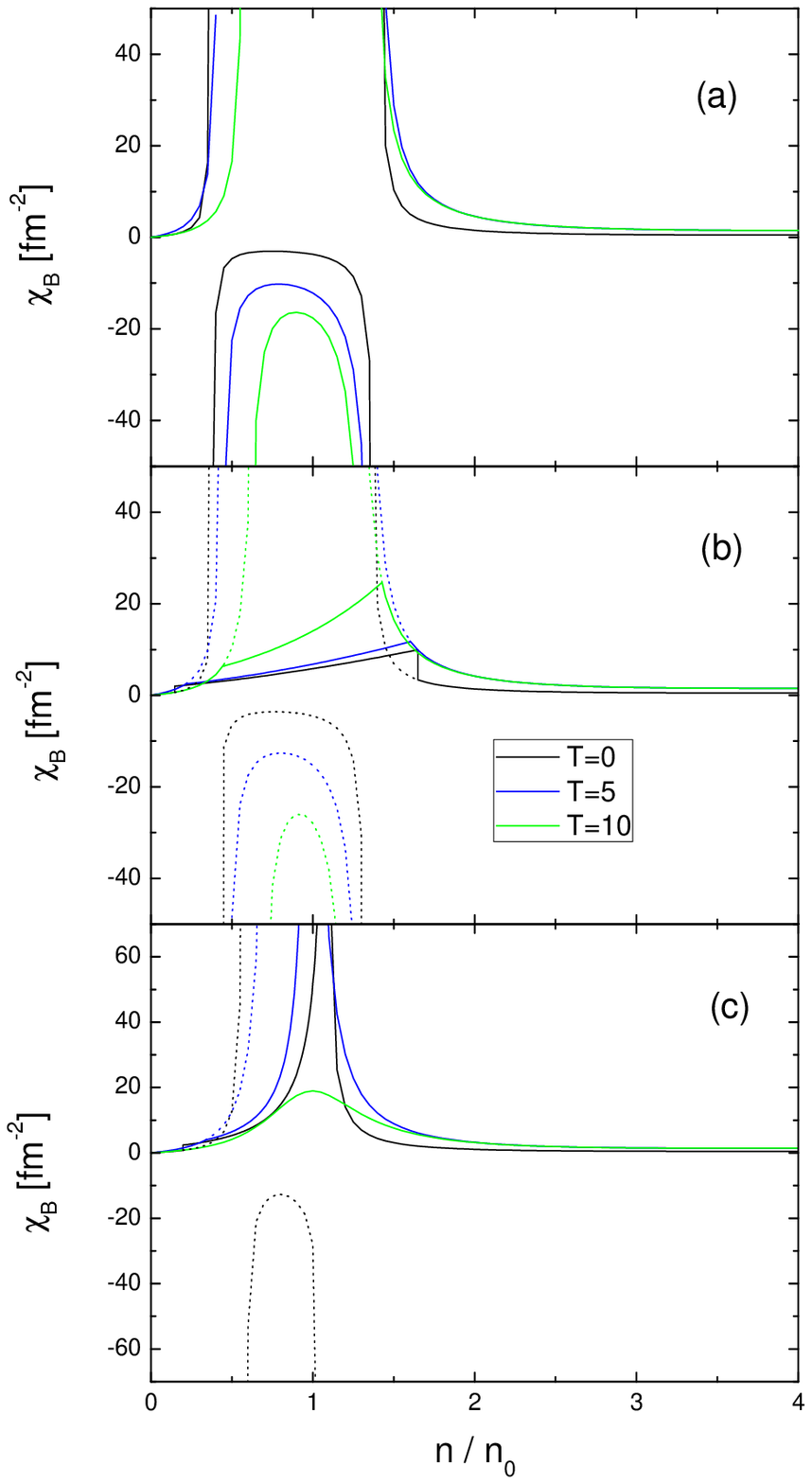}\vspace{-0.5cm}
    \caption{The second order susceptibility corresponding to the baryonic number as function
    of the quark number density for several temperatures and $x=0$ (a), $x=1/3$ (b), and $x=2/3$ (c).
    For $x=0$ only the results using the unstable equation of state are shown. In the panels b and c the solid lines
    correspond to the result including the ECR, dotted lines represent the result with contributions from
    unstable and metastable configurations.
    The reference density is $n_0=0.15$ fm$^{-3}$.}
    \label{Fig SuxB}%
\end{figure}
\begin{figure}
    \centering
    \includegraphics[height=0.4\textheight]{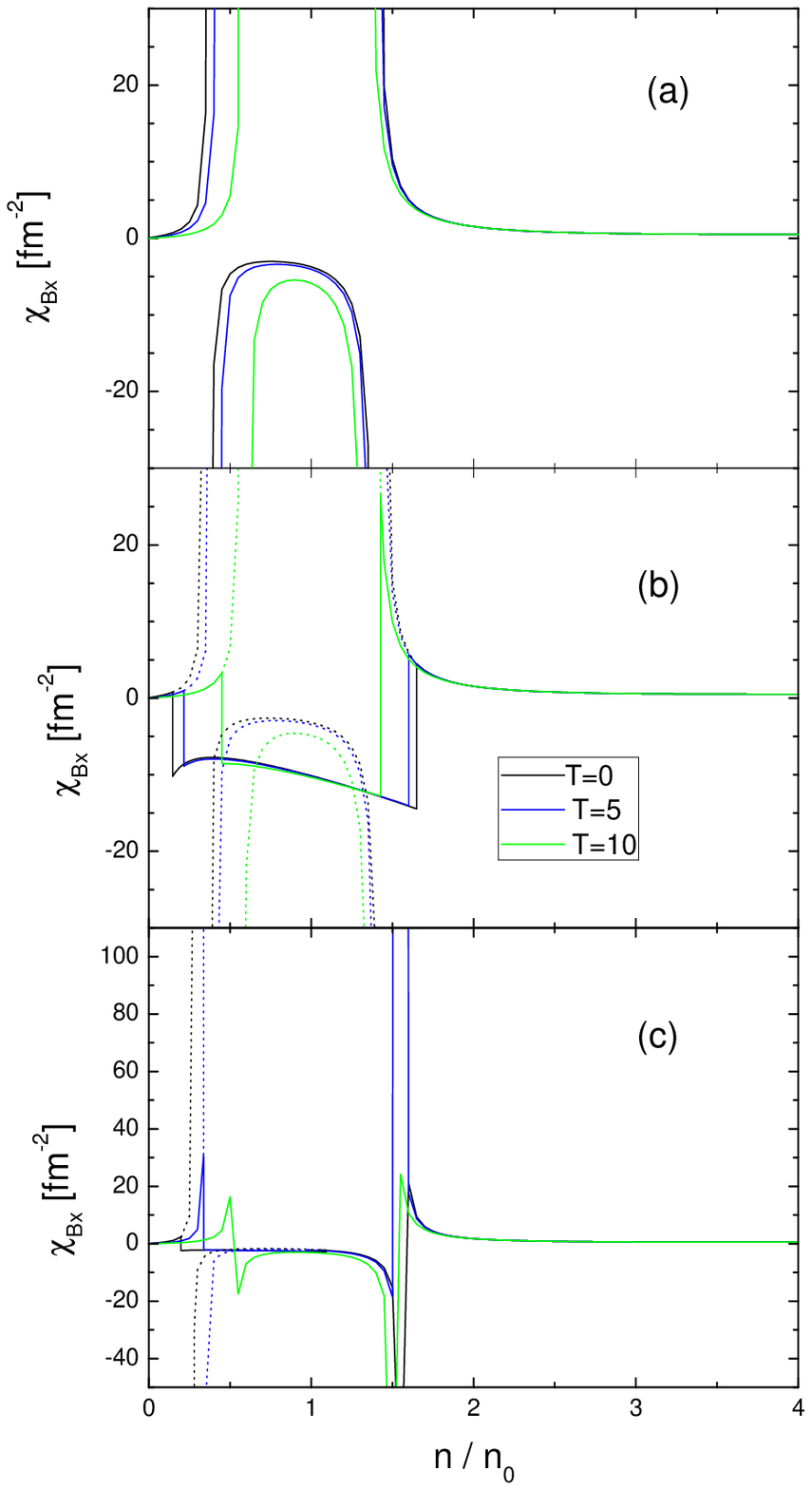}
    \caption{The second order susceptibility corresponding to the baryonic
    number (keeping constant the isospin fraction) as function of the quark number
    density for several temperatures and $x=0$ (a), $x=1/3$ (b), and $x=2/3$ (c). For $x=0$
    only the results using the unstable equation of state are
    shown. In the panels b and c the solid  lines
    correspond to the result including the ECR, dotted lines represent the result with contributions from
    unstable and metastable configurations.
    The reference density is $n_0=0.15$ fm$^{-3}$.}
    \label{Fig SuxBX}%
\end{figure}

The effects of the chiral transition on the number
susceptibilities have been discussed previously within the NJL
model \cite{SASAKI,SASAKI1,SASAKI2,COSTA2}, and also within the
LSMq \cite{STOKIC,SKOKOV,SKOKOV1,SKOKOV2,ALMASI}. The presence of
states out of thermodynamical equilibrium has been particularly
analyzed in \cite{SASAKI1,SASAKI2}. Different definitions of the
susceptibilities are shown in the following two figures, as
functions of the quark density for several temperatures within the
range of the first order transition. For the susceptibilities
associated with the baryon number, $\chi_B$ (Fig.\ref{Fig SuxB})
and $\chi_{Bx}$ (Fig.\ref{Fig SuxBX}), strong oscillations are
found at the borders of the ECR. For higher densities all the
curves decay and reach quickly the asymptotic regime. In fact
$\chi_B$ and $\chi_{Bx}$ coincide for $w=0, \, T=0$.

For $x=0$ (Figs. \ref{Fig SuxB}a, and \ref{Fig SuxBX}a) all the
curves diverge at the boundaries of the unstable section of the
equation of state, taking opposite signs on both sides. The same
behavior was  reported and discussed in \cite{SASAKI1,SASAKI2} for
calculations using the NJL model.

For $x>0$ the contributions of the unstable region are
distinguished by using dashed lines, while the results coming from
the ECR are represented by solid lines. The effect of the Gibbs
procedure is clear for $x=1/3$ (Figs. \ref{Fig SuxB}b and \ref{Fig
SuxBX}b), where the divergences are replaced by smooth curves with
finite discontinuities at the end points. In between $\chi_B$
remains positive and increasing with the density, a fact that is
emphasized as the temperature grows. On the contrary, $\chi_{Bx}$
experiences a drop in the ECR and seems insensitive to the
temperature.

Qualitative changes occur for the higher asymmetry $x=2/3$. For
$T=10$ MeV the system is above the CEP  and $\chi_B$ shows a
regular behavior, while $\chi_{Bx}$ still shows two peaks as a
remainder of the active thermodynamical fluctuations. For lower
temperatures  $T=0, \text{and} \; 5$ MeV, both susceptibilities
have large but finite discontinuities within the ECR. However it
must be bear in mind that this is an extrapolation, since the
effect of the weak forces will not be negligible for $x=2/3$.

\begin{figure}
    \centering
    \includegraphics[height=0.4\textheight]{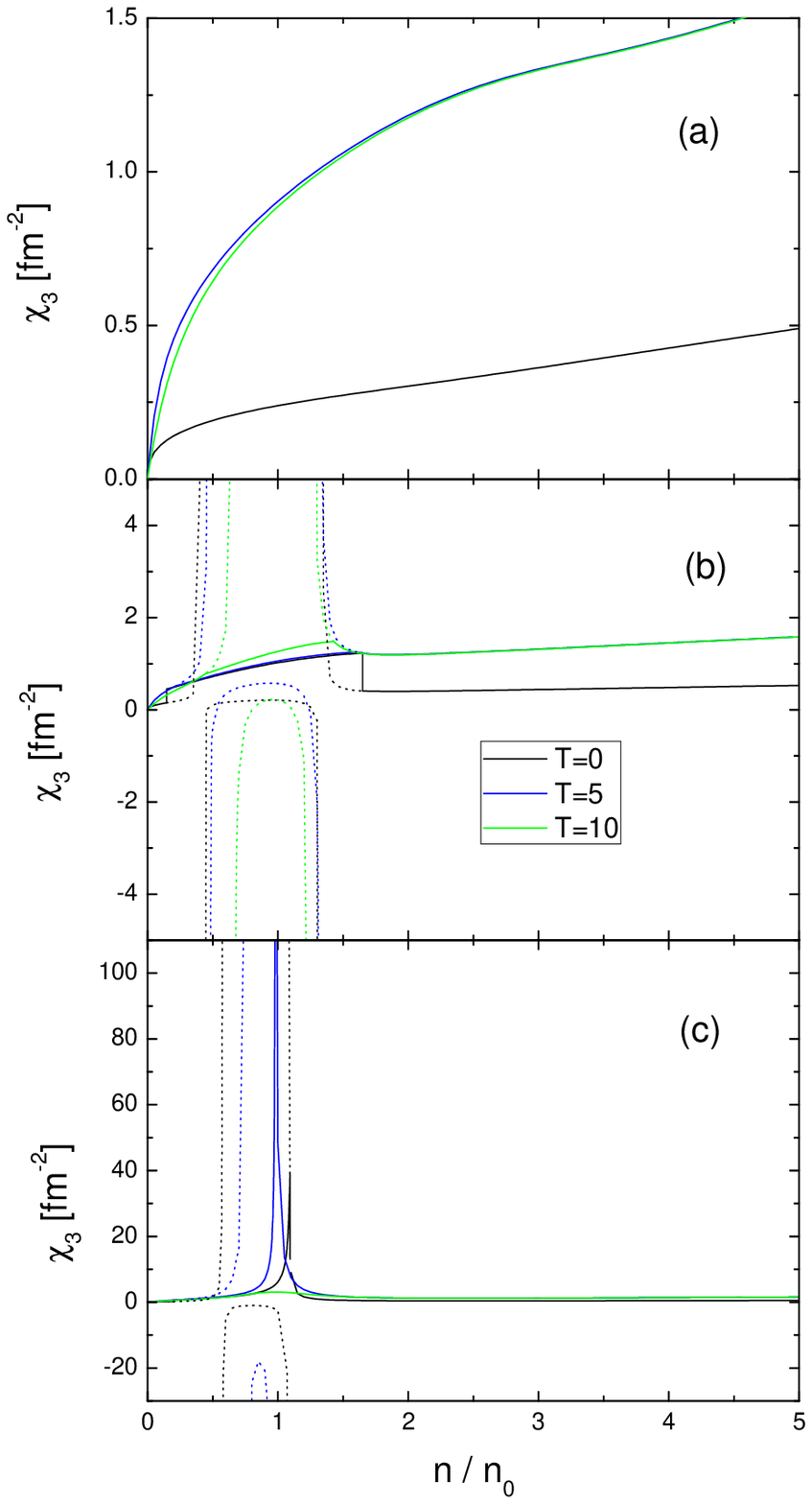}
    \caption{The second order susceptibility corresponding to the isospin number as function
    of the quark number density for several temperatures and $x=0$ (a), $x=1/3$ (b), and $x=2/3$ (c).
    For $x=0$ only the results using the unstable equation of state are shown. In the panels b and c the solid  lines
    correspond to the result including the ECR, dotted lines represent the result with contributions from
    unstable and metastable configurations. The reference density is $n_0=0.15$ fm$^{-3}$.}
    \label{Fig Sux3}%
\end{figure}

\begin{figure}
    \centering
    \includegraphics[height=0.4\textheight]{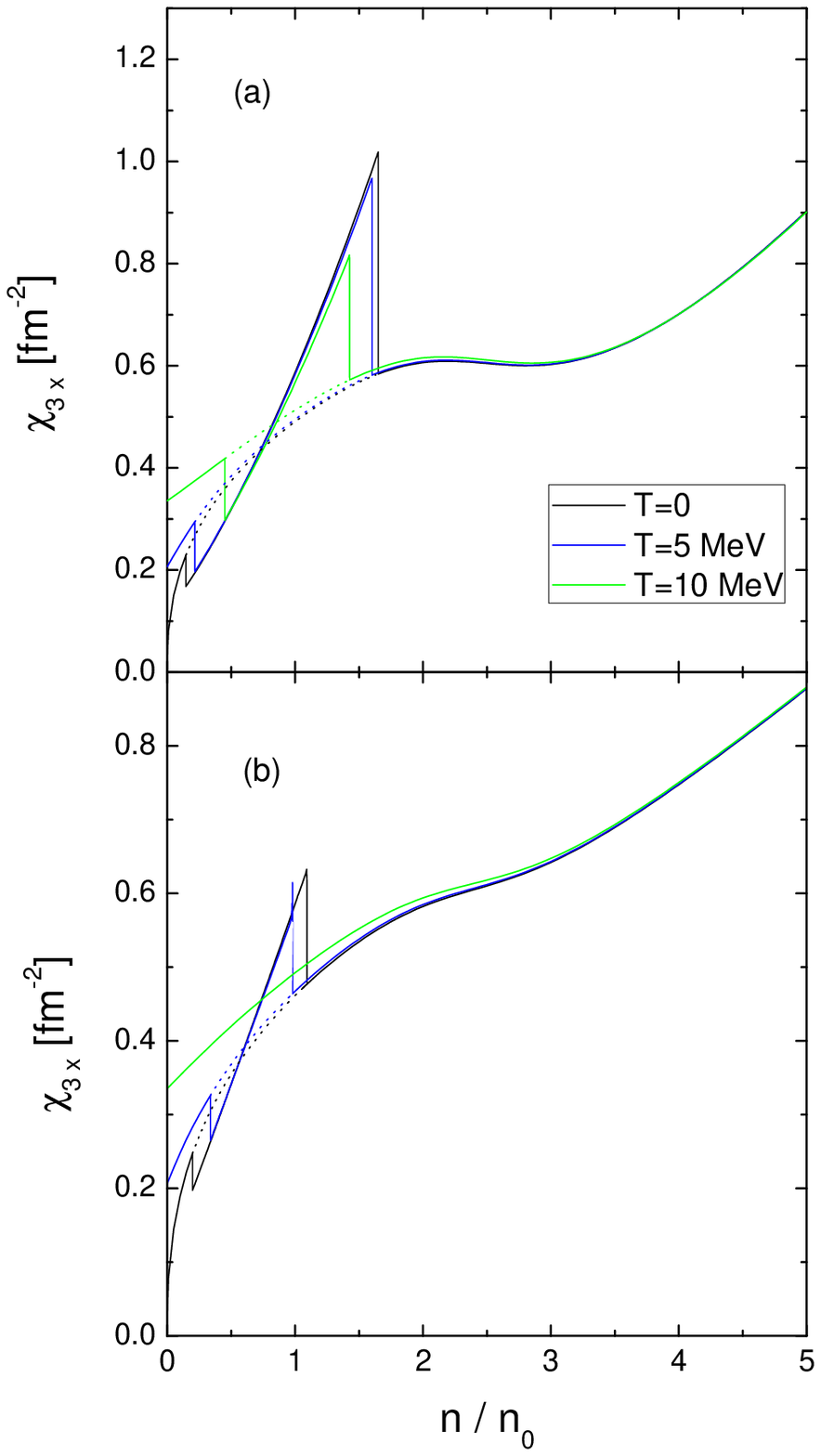}
    \caption{The second order susceptibility corresponding to the isospin number
    (keeping constant the isospin fraction) as function of the quark number density
    for several temperatures and $x=1/3$ (a), and $x=2/3$ (b). Solid lines
    correspond to the result including the ECR, dotted lines represent the result with contributions from
    unstable and metastable configurations. The reference density is
    $n_0=0.15$ fm$^{-3}$.}
    \label{Fig Sux3X}%
\end{figure}

The susceptibilities associated with the isospin composition are
presented in Figs. \ref{Fig Sux3} and \ref{Fig Sux3X} as functions
of the quark number density and several temperatures. The behavior
of $\chi_3$ for symmetric matter is completely regular despite the
presence of the binodal. The same applies to $\chi_{3x}$ for all
the cases shown in Fig. \ref{Fig Sux3X}, and is the transition
through the ECR the responsible of the discontinuities and an
almost linear increase with the quark number, independently of the
temperature.


\begin{figure}[b]
    \centering
    \includegraphics[height=0.4\textheight]{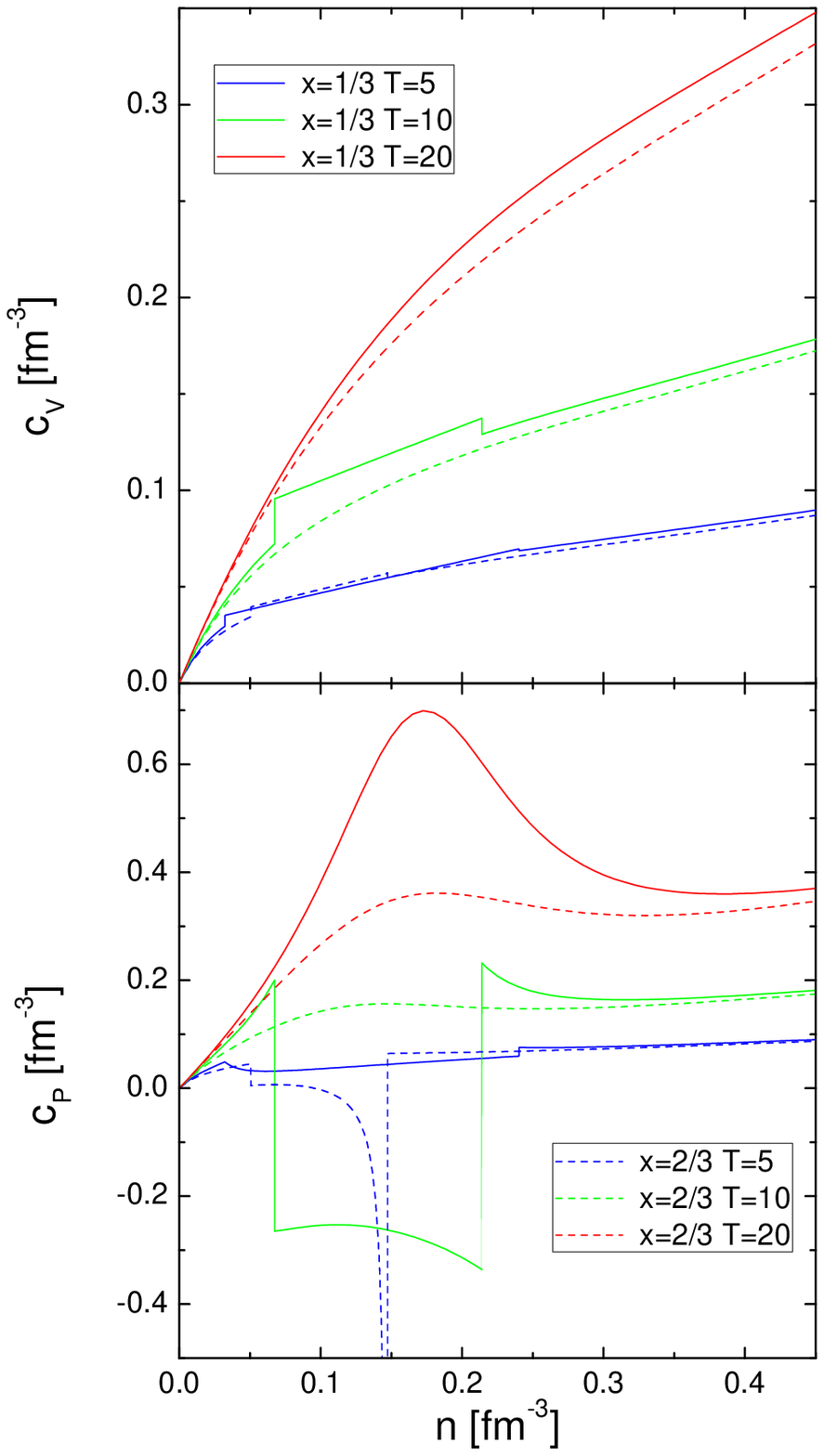}
    \caption{The specific heat at constant volume (a) and at constant temperature
    (b) as functions of the quark number density for the asymmetries $x=1/3, 2/3$,
    and temperatures $T=5, 10, 20$ MeV. The line convention is presented in two
    separate tables, one on each panel. }
    \label{Fig Heat}%
\end{figure}

As the next step the specific heat capacities are considered in
Fig.\ref{Fig Heat}. Both the constant volume and isobar quantities
are displayed for isospin asymmetric matter, showing the results
coming from the ECR only. The smooth variation of $c_V$ within the
binodal was discussed in \cite{SASAKI2}. In the present case
finite discontinuities are found at the extremes of the ECR which
do not affect significantly the general increasing trend. The
isobaric counterpart $c_P$ exhibits a more interesting behavior, a
strong dependence on $x$ is shown either in the crossover regime
($T=20$ MeV) or near the CEP ($T=10$ MeV). In the last case a drop
to $c_P<0$ occurs within the ECR. Such unexpected result has also
been found in nuclear physics and considered as a signal of the
liquid-gas transition \cite{CHOMAZ,BORDERIE,DE}. The curves for
$x=1/3$ show drastic thermal effects around the CEP, as it is
evident by comparing the $T=20$ MeV and $T=10$ MeV cases. This
effect becomes extreme for the higher asymmetry $x=2/3$ and
temperatures $T=5-10$ MeV, passing from a smooth dependence to a
negative peak concentrated at the border of the ECR.

\section{SUMMARY AND CONCLUSIONS}\label{SUMMARY}

In this work the chiral phase transition of homogeneous quark
matter, assuming the conservation of both baryonic and isospin
numbers, has been studied. This subject has been intensively
studied in recent years, motivated in part by the availability of
LQCD calculations, which overcame technical difficulties in the
limit of zero baryon density. Thus, investigations using effective
models have found a source of reliable data to contrast with their
predictions and an opportunity to improve their frameworks. Since
LQCD is efficient in the limit of zero baryonic density, much of
the recent theoretical efforts have scrutinized such regime
\cite{SON,STIELE,AVANCINI,LOPES,ADHIKARI,AYALA,AYALA0}. The chiral
transition at non-zero baryonic density is still the domain of
theoretical speculation by using effective models. Interesting
effects have been pointed out, such as the manifestations of
thermodynamical fluctuations near the non-equilibrium region of
the first order phase transition
\cite{SASAKI,SASAKI1,SASAKI2,STOKIC,ALMASI}. By definition this
domain can not be analyzed by the equilibrium thermodynamics.
However, if multiple charges are conserved and a low surface
tension is assumed, the system could undergo a continuous
transition through the coexistence of equilibrium states
\cite{GLENDENNING,HEISELBERG}. The LSMq with two flavors has been
used here to explore this hypothesis, introducing the scalar
isovector meson $\zeta$ as a necessary element to take account of
isospin imbalanced quark matter at finite density. The
calculations have been performed in the MFA with vacuum
contributions of the quark sector. The analysis of the effects of
higher order mesonic corrections to this scheme is in progress.
Neither pion condensation nor quark superconductivity have been
considered since the chemical potentials $\mu_B,\; \mu_3$ are
below the corresponding thresholds for the range of this
calculation.

The assumptions made in the present approach, as well as the range
of variables considered, make these results more appropriate for
astrophysical situations, as for instance the structure and
dynamics of self-bound stars \cite{ZACCHI}.

 The phase diagram obtained here is in
qualitative agreement with present knowledge, a smooth crossover
at high temperatures ends at a CEP, and changes to a first order
transition at low temperatures. Although the value obtained for
the critical temperature is too low, it is expected that higher
order mesonic corrections will improve this result
\cite{BILIC,PETROPOULOS,AYALA1}. The transition lines for several
isospin asymmetries have been obtained, showing that along the
crossover the transition temperature decreases slightly with
increasing the isospin asymmetry $x$. The first-order transition
occurs as a continuous succession of equilibrium states obtained
by the Gibbs construction.

The equation of state as well as  several bulk properties of quark
matter exhibit important modifications through the ECR. In the
crossover regime, the isothermal speed of sound at constant $x>0$
has a non-monotonic dependence on the particle density. It
presents a minimum in coincidence with the transition point.
Within the coexistence region the speed $v_T$ experiences a
noticeable decrease with finite discontinuities at the borders. A
mild variation of $v_T$ with density and temperature is predicted
for this domain.

The susceptibility associated with the baryonic number, both using
the standard definition ($\chi_B$) as well as that keeping $x$
constant ($\chi_{B x}$), remain bounded along the ECR for low
asymmetry parameter $x>0$. This result contrast with calculations
using the unstable sector of the equation of state. A distinctive
difference is that $\chi_B$ remains positive inside the ECR. As
$x$ grows, large but finite discontinuities appear at the high
density border of the ECR.

A similar description applies to $\chi_3$, with the noticeable
difference that for $x=0$ it remains bounded as predicted in
previous calculations \cite{SASAKI}. However the presence of the
binodal has clear manifestations for $x>0$.

In regard of the specific heat $c_V$, only small deviations from
the continuous behavior are introduced by the ECR. The isobaric
counterpart $c_P$ takes unexpected negative values, a fact known
in nuclear physics and interpreted as a signal of the spinodal
fragmentation \cite{CHOMAZ,BORDERIE,DE}.

\section*{Acknowledgements}
This work has been partially supported by CONICET, Argentina under
the project 11220200102081CO, and by UNLP, Argentina.

\appendix
\section{}

The vacuum contribution to the grand potential has been discussed
in numerous works. In particular it has been pointed out
\cite{SKOKOV}, that its inclusion in the two flavor LSMq changes
the character of the chiral transition. For the sake of
completeness, the deduction of Eq.(\ref{GP VAC}) is shown in the
following.

The contribution of the Dirac sea to the thermodynamical
potential, see for instance \cite{DOLAN}, is given by
\begin{equation} \omega_{\text{vac}}=-2 N_c \sum_{j=u,d}\int \frac{d^3p}{(2
\pi)^3} E_j,\label{App1}\end{equation}
where $E_j^2=p^2+m_j^2$, and the integral is clearly ultraviolet
divergent. However, the divergence can be expressed as an isolated
pole using dimensional regularization. For this purpose the
integral is rewritten in terms of the causal Dirac propagator as
\[I_j=\int \frac{d^3p}{(2\pi)^3} E_j=2 i \int
\frac{d^4p}{(2\pi)^4}\,\frac{p_0^2}{p_0^2-E_j^2+i \epsilon}=
\lim_{\varepsilon \rightarrow 0} 2 i \int
\frac{d^dp}{(2\pi)^d}\,\frac{p_\mu p^\nu}{(p_0^2-E_j^2)^\alpha}.
\]
On the right hand side the integration is carried out in a space
of dimension $d=4-2 \varepsilon$, and $\mu=\nu=0, \, \alpha=1$.
This is a typical expression which can be found in several texts,
leading to the result
\[I_j=\left(\frac{m^2_j}{4 \pi} \right)^2\lim_{\varepsilon \rightarrow 0}
\left(m^2_j\right)^{-\varepsilon}\,
\Gamma(\varepsilon-2)=\frac{m^4_j}{32 \pi^2} \,\lim_{\varepsilon
\rightarrow 0}
\left[\frac{1}{\varepsilon}+\frac{3}{2}-\gamma-\ln(m_j^2)+\mathcal{O}(\varepsilon)\right].
\]
In the right hand side the divergence appears as a simple pole at
$\varepsilon=0$ with main coefficient $m^2_j/32 \pi^2$. For this
reason one can define a regularized integral by taking the
reference state with zero density and temperature, where the quark
mass is degenerate $m_j=m_0$, and applying a subtraction scheme
\begin{eqnarray}I_j^{\text{reg}}&=&I_j-I_j(m_j=m_0)-\left(m_j^4-m_0^4\right)\,\frac{d
I_j}{dy}(m_j=m_0)\nonumber \\
&=&\frac{1}{64
\pi^2}\left[m_j^4-m_0^4-m^4_j\,\ln\left(\frac{m^4_j}{m^4_0}\right)\right]
\nonumber\end{eqnarray} where $y=m_j^4$. Thus the regularized
contribution from the quark Dirac sea is
\begin{equation}\omega_{\text{vac}}=\frac{N_c}{32
\pi^2}\sum_{j=u,d}\left[m_0^4-m_j^4+m^4_j\,\ln\left(\frac{m^4_j}{m^4_0}\right)\right]\label{App2}\end{equation}

\bibliography{PRC}

\end{document}